\documentclass[prb,twocolumn,showpacs,amsmath,amssymb,amsfont,floatfix]{revtex4}

\usepackage{graphicx}

\begin{document}

\title{Observation of double resistance anomalies and excessive resistance 
in mesoscopic superconducting Au$_{0.7}$In$_{0.3}$ rings with phase separation}

\author{H. Wang}

\author{M.~M. Rosario}
  \altaffiliation[Present address: ]{Department of Physics and Astronomy,
Saint Mary's College of California, Moraga, CA 94556}

\author{H.~L. Russell}

\author{Y. Liu}

\affiliation{Department of Physics, The Pennsylvania State University,
University Park, PA 16802}

\date{\today}

\begin{abstract}

We have measured mesoscopic superconducting Au$_{0.7}$In$_{0.3}$ rings
prepared by e-beam lithography and sequential deposition of Au and In
at room temperature followed by a standard lift-off procedure. In samples
showing no Little-Parks resistance oscillations, highly unusual double
resistance anomalies, two resistance peaks found near
the onset of superconductivity, were observed. Although resistance anomaly
featuring a single resistance peak has
been seen in various mesoscopic superconducting samples, double resistance
anomalies have never been observed previously. The dynamical resistance
measurements suggest that there are two critical
currents in these samples. In addition, the two resistance peaks were found to be
suppressed at different magnetic fields. We attribute the observed double
resistance anomalies to an underlying phase separation in which In-rich grains of
intermetallic compound of AuIn precipitate in a uniform
In-dilute matrix of Au$_{0.9}$In$_{0.1}$. The intrinsic
superconducting transition temperature of the In-rich grains is substantially
higher than that of the In-dilute matrix. The suppression of the
conventional Little-Parks resistance oscillation is explained in the same
picture by taking into consideration a strong variation in the $T_c$ of
the In-rich grains. We also report the observation of an unusual 
magnetic-field-induced metallic state with its resistance higher
than the normal-state resistance, referred to here as excessive resistance, 
and an h/2e resistance oscillation with the amplitude
of oscillation depends extremely weakly on temperature.

\end{abstract}


\maketitle


Recent structural and electrical transport studies of Au$_{0.7}$In$_{0.3}$
films revealed an interesting phase separation in which In-rich grains,
most likely intermetallic compound of AuIn, precipitate in a uniform
In-dilute matrix, most likely Au$_{0.9}$In$_{0.1}$, with the superconducting
transition temperature of the In-rich grains substantially higher than that of
the In-dilute matrix, forming an array of 
superconductor-normal metal-superconductor (SNS) Josephson 
junctions~\cite{zador02}, with both the local
gap and Josephson coupling between adjacent In-rich grains varying randomly. These films
represent a novel system for studying the disorder effects on superconductivity
in two dimension (2D).  Conventional granular films, such as those
prepared by quench deposition, on the other hand, are modeled as a random array
of superconductor-insulator-superconductor (SIS) Josephson junctions. The phase
of the superconducting order parameter, $\phi$, and the number of Cooper pairs,
$N$, are conjugate variables quantum-mechanically, and are subject to an
uncertainty relation. Therefore, the confinement of Cooper pairs in SIS granular
films due to charging energy leads to the phase fluctuation, and a 2D
superconductor-to-insulator transition (SIT) if the phase fluctuation is
sufficiently strong. Cooper pairs in Au$_{0.7}$In$_{0.3}$ films, however,
are not subject to similar spatial confinement. Nevertheless, a quantum
superconductor-normal metal phase transition (SNT) is still expected in such a
random array of SNS Josephson junctions, as predicted theoretically recently
~\cite{feige01,spiva01}.

Experimentally interesting physical phenomena have been found in
planar and cylindrical films of
Au$_{0.7}$In$_{0.3}$~\cite{zador01_1,zador01_2}. In particular,
an h/4e, rather than h/2e, resistance oscillation, was found~\cite{zador01_2}.
While the physical origin of this h/4e resistance oscillation 
is not fully understood, it may be associated with the
presence of $\pi$-junctions between adjacent In-rich grains that possess a
negative rather than positive Josephson coupling constant~\cite{zador01_2}. Such negative
Josephson coupling could be derived from mesoscopic fluctuation or correlation
effects~\cite{spiva91,kivel92}. The presence of these random distributed
$\pi$-junctions leads to an h/4e resistance oscillation because of the
ensemble average~\cite{kivel92}. In single mesoscopic rings of
Au$_{0.7}$In$_{0.3}$, the
resistance oscillation is expected to be of the conventional h/2e period, but
with an unconventional phase shift by $\pi$ if the number of $\pi$-junctions
in the ring is odd. No phase shift is expected for rings with even number of
$\pi$-junctions.

In this paper, we report our experimental studies of mesoscopic superconducting
Au$_{0.7}$In$_{0.3}$ rings prepared by e-beam lithography and sequential
deposition of Au and In at room temperature followed by a standard
lift-off procedure. Conventional h/2e resistance oscillation was observed,
as expected. However, whether there is a phase shift by $\pi$ in the h/2e
oscillation is not determined because of technical issues (see below). On
the other hand, in samples where the conventional Little-Parks (L-P) resistance
oscillation was suppressed, double resistance anomalies and a 
magnetic-field-induced metallic state with excessive resistance were found. We attribute
these observations to the separation of In-rich and In-dilute phases
in these rings.

\begin{figure}
\includegraphics{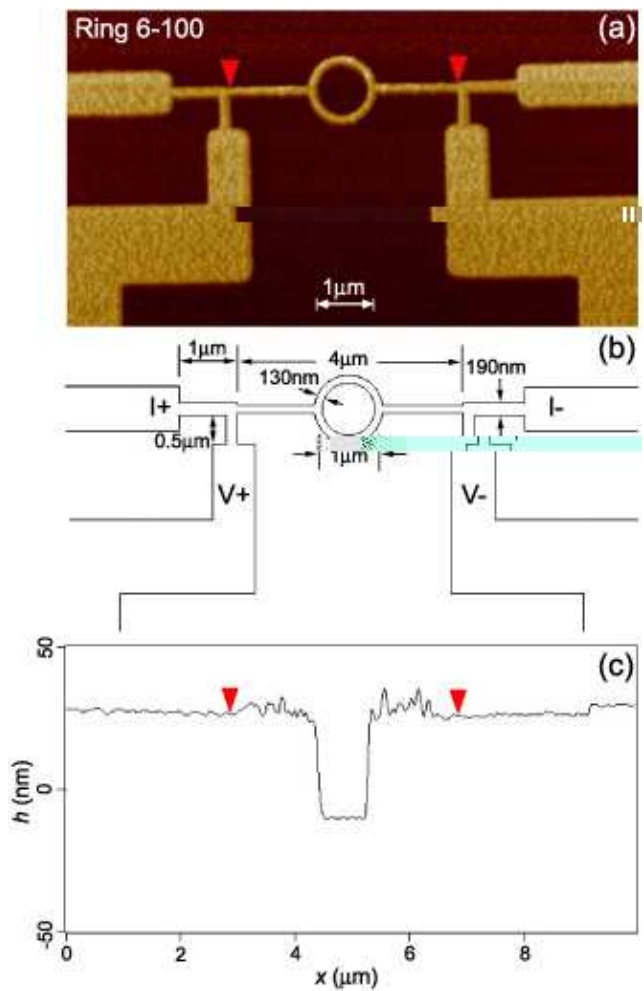}
\caption{\label{fig1} a) Atomic force microscope (AFM) image of a 
1-$\mu $m-diam
Au$_{0.7}$In$_{0.3}$ ring (6-100, see Table I); b) Schematic 
corresponding to the
AFM image in Fig. 1a; c) Height profile along the ring arms showing 
surface roughness.}
\end{figure}

Conventional e-beam lithography was used to prepare the Au$_{0.7}$In$_{0.3}$
rings. The pattern of several rings were generated using double-layer
PMMA/MMA resist on 
polished 1cm$\times $1cm sapphire substrate. Sequential thermal evaporation
of alternating 99.9999\% pure Au and In layers, with the layer 
thickness determined by the appropriate atomic ratio of Au to In, was 
carried out at
ambient temperature in a conventional evaporator with a vacuum of $1\times
10^{-6}$ torr or slightly better. The ring pattern was placed with respect
to the Au and In sources so as to minimize the shadow effects during
evaporation. Atomic force microscope (AFM) was used to image the resulted
Au$_{0.7}$In$_{0.3}$ rings before and/or after the measurements.
The electrical transport measurements were carried out in a dilution fridge
which is equipped with a superconducting magnet and has a base temperature $<$20mK. 
All electrical leads entering the cryostat were
filtered by RF filters
with an attenuation of 10dB at 10MHz and 50dB at 300MHz.
Resistance characteristics were measured with a d.c. current source
and a nanovoltmeter. The magnetic
field was applied perpendicular to the substrate, the plane of the rings.

\begin{figure}
\includegraphics{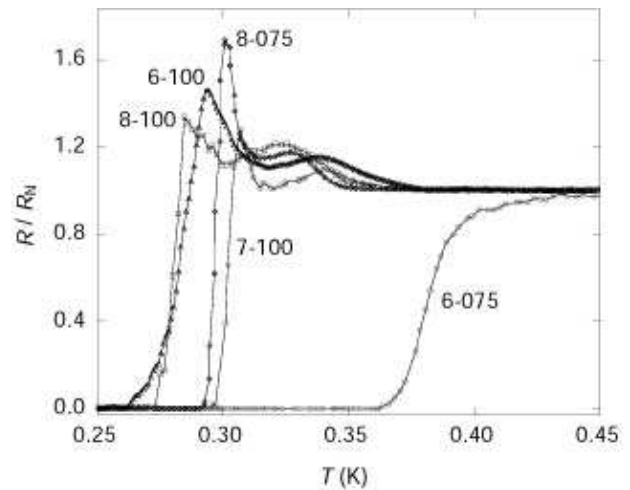}
\caption{\label{fig2} Normalized resistances as function of temperature
$R(T)$ for five Au$_{0.7}$In$_{0.3}$ rings. Rings are labeled as indicated.
Parameters for these rings are shown in Table I.}
\end{figure}

Five rings were measured in this study. Two rings have a diameter of
0.75$\mu $m and the rest have a diameter of 1$\mu $m. The thickness
of all rings is nominally 30 nm. An AFM image of a 1-$\mu $m-diam ring, Sample 6-100,
is shown in Fig. 1a, with a schematic shown in Fig. 1b to detail the
various dimensions of the sample. Rings with a 0.75-$\mu $m diameter have the
same layout with a 4 $\mu $m voltage-probe separation. Two features of
the samples should be noted. First, as seen in the schematic of the sample
  (Fig. 1b), the linewidth increases slightly at the nodes of the
voltage leads going towards the large contact pads. Such variation of the
linewidth, observed in all samples, was probably due to an overexposure
while writing the large contact pads; Second, the surface roughness as
seen by AFM increases at the narrow part of the sample (within the voltage
leads, see Fig. 1c). However, the cause for the increased height variations
is not clear. We found in a separate study that the lift-off process we used
for Au$_{0.7}$In$_{0.3}$ rings could generate rough edges (due to resist
residue that is not lifted). Therefore the surface roughness may
not necessarily indicate that the Au$_{0.7}$In$_{0.3}$ rings themselves have
a rough surface.

In Fig.~2, normalized resistance as function of temperature under zero magnetic
field is shown for all five rings. The structural and electrical parameters
are listed in Table I. The typical values for the normal-state resistivity, $\rho_N$, are about
3 times larger than that for Au$_{0.7}$In$_{0.3}$ films with the same 
thickness~\cite{zador02}, which is reasonable because of the extra 
surface
scattering encountered in mesoscopic samples.

\begin{table}
\caption{\label{tab1} Selected parameters for the
rings. $T_{\rm c}$ is determined at the onset of the resistance 
drop for Ring 6-075, and for the rest of rings at the onset of
the high-temperature resistance peak (HTRP). $d$ is diameter.
$t$ is thickness.}
\begin{ruledtabular}
\begin{tabular}{cccccc}
   Sample & $d$ ($\mu $m) & $t$ (nm) & $R_{{\rm N},\Box}$ ($\Omega $)
& $\rho_N$ ($\mu \Omega $ cm) & $T_{\rm c}$ (K)\\ \hline
   6-075 & 0.75 & 30 & 13.5 & 40.6 & 0.450\\
   6-100 & 1.00 & 30 & 13.5 & 40.4 & 0.380\\
   7-100 & 1.00 & 30 & 12.5 & 37.4 & 0.370\\
   8-075 & 0.75 & 30 & 11.3 & 33.8 & 0.359\\
   8-100 & 1.00 & 30 & 12.4 & 37.2 & 0.355\\
\end{tabular}
\end{ruledtabular}
\end{table}

From Fig.~2 it is clear that there exist two types of behaviors among these
rings. For Ring 6-075, a smooth resistive transition was seen. For other
four rings, however, two resistance peaks were found near the onset of the
superconducting transition. It is seen that the low-temperature resistance 
peak (LTRP) is relatively sharp, with a resistance
about 30-70\% higher than the normal-state resistance, $R_{\rm N}$. The
high-temperature resistance peak (HTRP) is broader
and smaller in height (10-20\% higher than $R_{\rm N}$) than LTRP.

\begin{figure}
\includegraphics{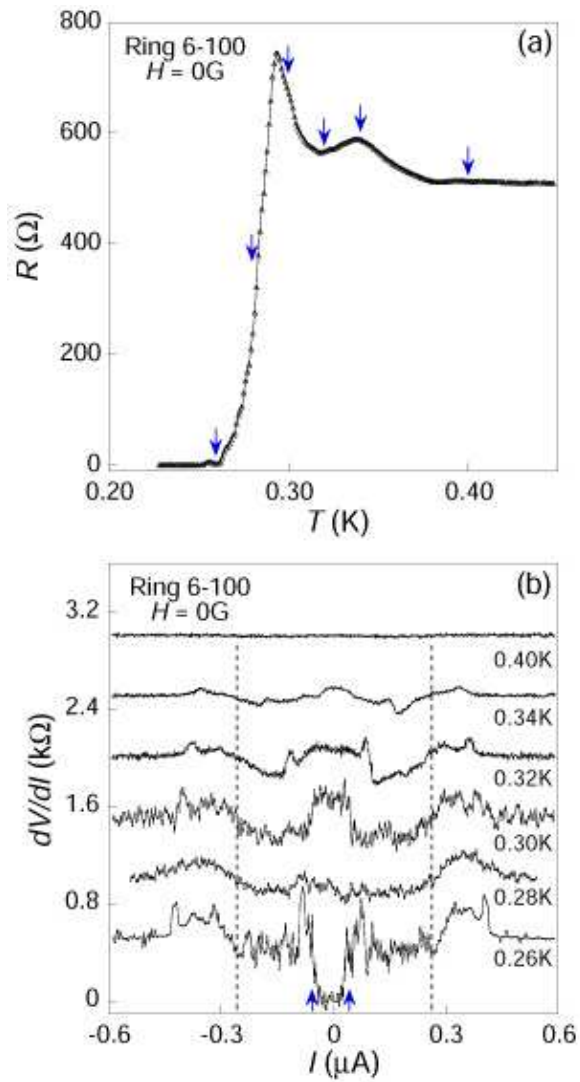}
\caption{\label{fig3} a) Normalized $R(T)$ for Ring 6-100 in zero field.
The temperatures at which $dV/dI$ were taken (shown in b) are indicated by
arrows; b) Differential resistance $dV/dI$ (calculated from d.c. current
biased $I-V$ curves using numerical derivatives) at various temperatures as
indicated. Two critical currents are indicated by the dashed line and by the
arrow. All curves except the one for $T = 0.26$K are shifted for clarity.}
\end{figure}

\begin{figure}
\includegraphics{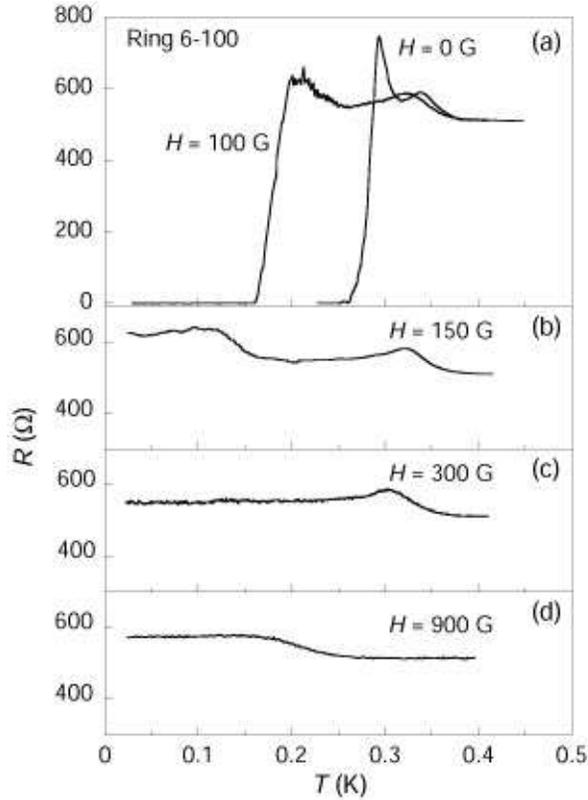}
\caption{\label{fig4} $R$($T$) at several magnetic fields as
indicated for Ring 6-100. The fields were applied
perpendicular to the plane of the ring. A resistance plateau
higher than $R_{\rm N}$ is shown down to the lowest temperature
available ($<$ 20mK) for $H = 300$ and $900$G.}
\end{figure}

The double resistance anomalies observed in
these four rings are highly unusual. Single resistance anomaly, a resistance
peak right at the onset of superconductivity, was previously observed in
mesoscopic structures of Al and other materials
~\cite{santh91, moshc94,park97,strun98,aruty99,landa97,moshc97}.
Temperature dependence of the dynamical resistance as a function of bias
current was also measured~\cite{santh91}. It seems reasonable that the double
resistance anomalies observed in the current work and the previously
observed single resistance anomaly share similar physical origin. Since the single resistance anomaly
was found to occur at the onset of superconductivity, the occurrence of the double resistance
peaks appears to indicate the existence of two
superconducting phases with slightly different $T_c$'s. Based on our previous
work on planar Au$_{0.7}$In$_{0.3}$ films~\cite{zador02}, these two phases should be 
the In-rich grains, most likely intermetallic compound of AuIn, and the uniform
In-dilute matrix, most likely Au$_{0.9}$In$_{0.1}$. The superconducting
transition temperatures of the In-rich grains are substantially higher than that of
the In-dilute matrix. As a result, the sample can be viewed as an array of 
SNS Josephson junctions. In this picture, HTRP corresponds to that
of the individual In-rich grains while LTRP is that of the junction array
formed by the In-rich grains and the In-dilute matrix.

In Fig.~3b, we plot the differential resistance $dV/dI$ as functions of the
bias current $I$ for Ring 6-100 at different temperatures of the resistive
transition as marked by arrows in Fig. 3a. At $T= 0.26$K, sharp rises in
$dV/dI$ were seen (shown by the dash lines and by arrows at 
0.26K), which appear to indicate the existence of two critical 
currents ($I_c$) in the sample. Essentially, the dynamical resistance
is vanishing at low bias currents, and rises sharply at $\approx 0.05\mu$A, the
lower critical current. The differential resistance minimum below the lower
critical current was quickly suppressed by the increasing temperature,
replaced by a central peak near zero bias current, indicating that
the lower critical current vanishes. The peak at larger critical
current was found to change only slightly with the
increasing temperature, disappearing only when the entire sample turns normal.

Results from the dynamical resistance measurements therefore supports the 
idea that the double resistance anomalies resulted from the phase separation. 
In fact, features found in the dynamical resistance in low
currents (below 0.2$\mu$A) and their temperature dependence, especially the sharp
rises above lower $I_c$ and the emergence of a central peak
near zero bias current, are consistent with those observed
in mesoscopic Al wires with single resistance anomaly~\cite{santh91},
suggesting that the smaller characteristic current is the critical
current of the overall sample, the SNS junction array. The larger critical
current would then correspond to that of the individual In-rich grains, which
should drop slowly with increasing temperature.

\begin{figure}
\includegraphics{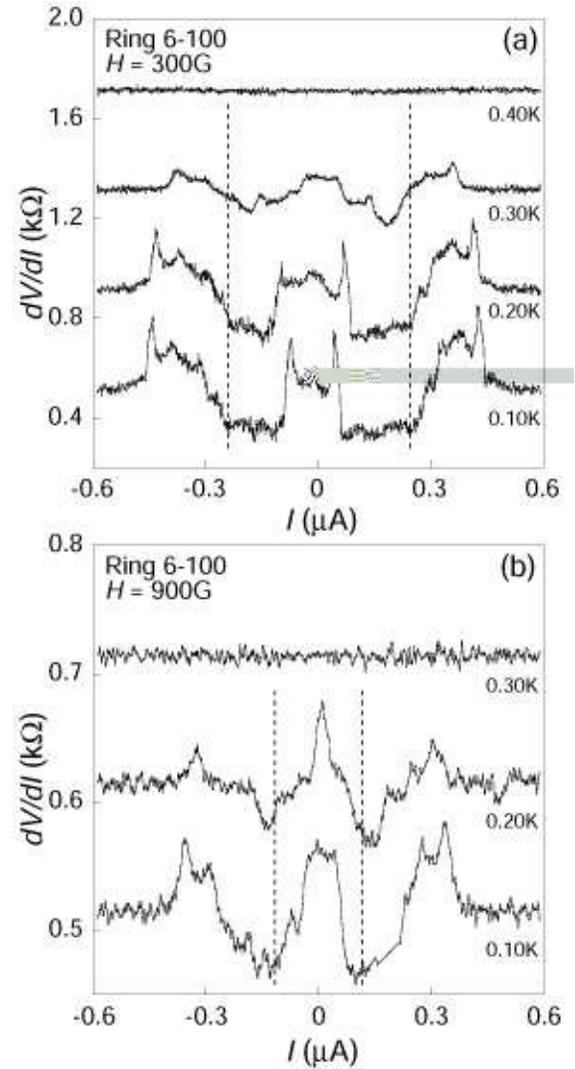}
\caption{\label{fig5} a) Differential resistance $dV/dI$ at various 
temperatures, as indicated, for Ring 6-100 at $H = 300G$; b) $dV/dI$ at
$H = 900$G. $dV/dI$ curves are based on
numerical derivatives of d.c. current biased $I-V$ measurements.
All curves except the one at $T = 0.10$K are shifted for clarity in both panels.}

\end{figure}

The measurements on the magnetic field dependence of the double resistance
anomalies, shown in Fig.~4, provided further support to the above picture.
It is seen that LTRP is affected significantly at a field as low as 100G,
becoming barely visible at 300G, which appears to be a critical field ($H_c$).
On the other hand, in a field as high as 900G,
even though HTRP is broadened and shifted to lower temperatures, it is
clearly visible. In fact, even at field up to 1300G, 
HTRP could still be identified close to the lowest temperature, 20mK.
It is therefore evident that the HTRP is associated with In-rich grains
with a critical field of 1300G while LTRP belongs to the SNS junction array
with a critical field of 300G.

The differential resistance curves taken at 300G (Fig.~5a) indeed show that,
at 0.1K, the minimum near the zero bias current is replaced by a peak, indicating that 
the critical current of the SNS junction array is essentially zero at 300G.
This is consistent with results of Fig. 4. This in turn supports
the assessment that LTRP is associated with the SNS junction array.
The sharp rise in $dV/dI$ at high bias current, on the other hand, is clearly visible
at $H = 900$G (Fig. 5b), suggesting that HTRP is associated with the
In-rich grains.

\begin{figure}
\includegraphics{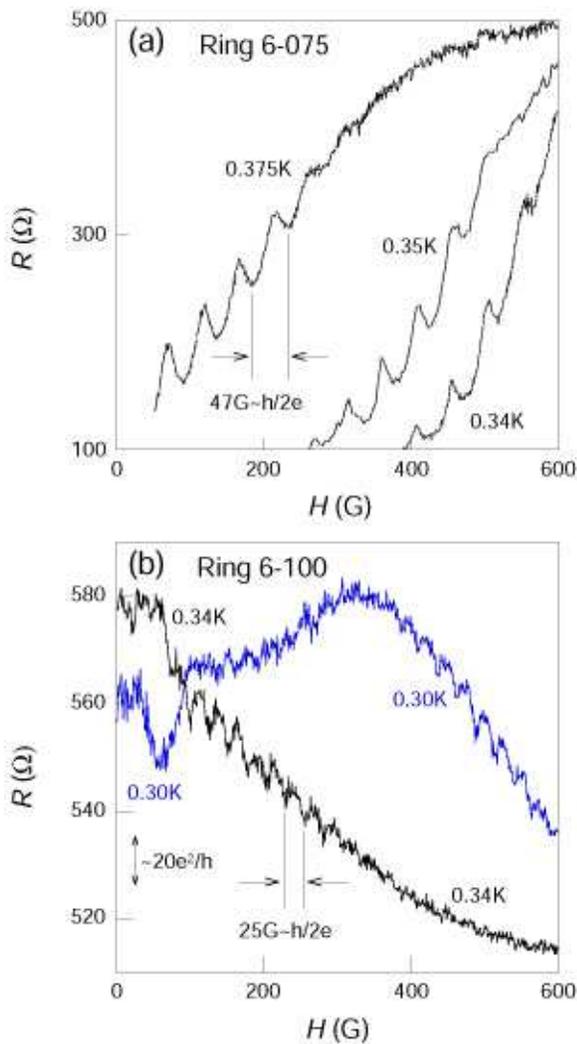}
\caption{\label{fig6} a) Little-Parks resistance oscillation of
Ring 6-075 at various temperatures as indicated. The period of
the oscillation corresponds to h/2e;
b) Magnetoresistance oscillations of Ring 6-100 at various temperatures
as indicated. The resistance oscillation of a period of h/2e is due to
coherent back scattering of single electrons. The resistance variation
corresponding to $20e^2/h$ is shown.}

\end{figure}

An interesting feature emerging from Fig.~4 is that, in the intermediate
field range, the broadened HTRP is seen to extended to the lowest temperature
we measured. At $H = 900$G, in particular, the resistance
became independent of temperature down to 20mK, with a resistance value
larger than that of the normal state $R_{\rm N}$. The existence of this
low-temperature resistance plateau, referred to here as excessive resistance,
appears to suggest the existence of a metallic state in which the In-rich
grains are superconducting, but not Josephson coupled. This metallic state cannot
be due to heating as the onset of the metallic behavior ($> 0.2$K) 
is too high for electrons to be at temperatures higher than the lattice. 
The observation of a metallic state with its resistance larger than the normal
state resistance and its onset temperature outside the low temperature region in which 
the heating is an issue, is itself a strong evidence that this metallic 
state is intrinsic. The physics of such a novel
metallic state is yet to be explored.

For Ring 6-075 with a smooth $R(T)$ showing no double resistance anomalies,
its onset $T_c$ is around 0.45K, close to that observed in thinnest 
planar films
(thickness $\leq 15$nm) of Au$_{0.7}$In$_{0.3}$~\cite{zador02}. In those
thinnest films, the interdiffusion of Au and In is suppressed by the
close proximity to the substrate, resulting in films that are uniform
rather than phase separated. Ring 6-075 may be of similarly uniform structure.
This assessment is consistent with the observation of Little-Parks (L-P) resistance oscillation
in this sample as shown in Fig.~6a, with a conventional 
period of h/2e. The observation of L-P resistance oscillation requires 
that the local $T_c$ to be reasonably close across the sample 
to ensure a rigid shift of the $R(T)$ curves with the
applied magnetic flux.

\begin{figure}
\includegraphics{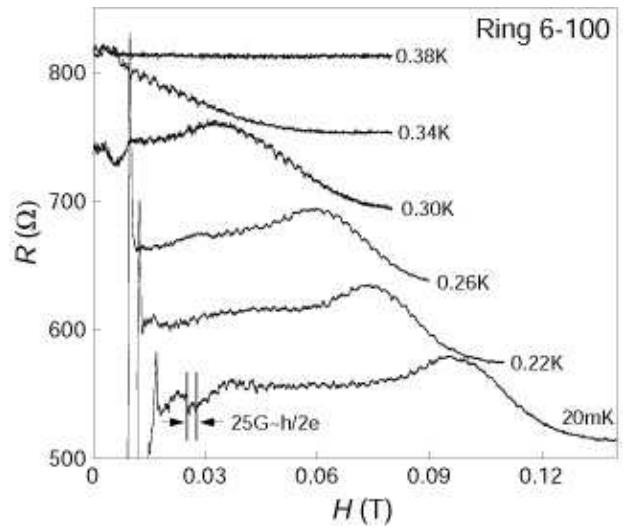}
\caption{\label{fig7} Resistance as a function of applied magnetic 
field $R(H)$ at
several temperatures, as indicated. The period of the resistance 
oscillations is
h/2e. Essentially no change of the oscillation amplitude is seen from 0.34K
down to 20mK. The negative magnetoresistance in the intermediate fields is
related to the suppression of the resistance anomalies by the field. Curves
are shifted for clarity except the one at 20mK.}

\end{figure}

We would like to mention that whether there was a phase shift by $\pi$ in
the h/2e oscillation in Ring 6-075, the question that motivated this work
originally, could not be determined. We found it difficult to determine the
flux sufficiently precisely because of the possibility of trapping flux in the
superconducting magnet.

For other four rings showing double resistance anomalies, the $T_{\rm c}$
of the In-rich grains will depend
sensitively on the atomic composition, the size, and the level of
intragrain disorder. Since the L-P resistance oscillation relies on
the rigid shift of $R(T)$ curve with the applied magnetic flux, a strongly
varying local $T_c$ will suppress the L-P resistance
oscillation. Indeed, experimentally, the L-P resistance oscillation
was not observed in the phase separated rings. Only a weak h/2e resistance oscillation was found
from near the transition temperatures down to the base temperature 20
mK (Fig. 6b). The amplitude of this oscillation is much smaller than 
the L-P oscillation in Ring 6-075 (Fig.~6a). 
As a result, this oscillation should be the Sharvin-Sharvin (S-S)
oscillation~\cite{sharv81}, predicted theoretically by Altshuler and Aronov~\cite{altsh81}.
This oscillation originates from the coherent backscattering
of normal electrons in disordered systems, a weak localization phenomenon.
In terms of conductance, $\Delta G \approx 11e^2/h$, where an effective
normal-state resistance of the ring itself, 105.8$\Omega$ (estimated
from the sample geometry), rather than the total resistance of the sample, 510$\Omega$, is used.
Clearly, the S-S oscillation is enhanced due to the existence of superconducting
In-rich grains in the sample~\cite{zador01_1, court96, petra93}. 
This also explains why S-S resistance oscillation
was observed only below the onset of superconductivity ($T = 0.38$K). Above
$T = 0.38$K, the dephasing length, $L_\varphi$, is smaller than the circumference
of the ring, $L = \pi d$. As the In-rich grains becomes superconducting, dephasing
occurs only in the In-dilute matrix, making the effective circumference
smaller than $L_\varphi$.
The associated negative magnetoresistance background shown in Fig.~6b is
related primarily to the suppression of the resistance anomalies by 
magnetic field.

While h/2e resistance oscillation observed in the 
phase separated rings is clearly S-S resistance oscillation, the 
temperature dependence
of the amplitude of the oscillation is puzzling. Over a wide range of
temperature, from 0.34K down to 20mK (Fig. 7), the amplitude of the
resistance oscillation is hardly changed. As the amplitude of
S-S oscillation depends on the dephasing length, and the
dephasing length increases with decreasing temperature, the resistance
oscillations should increase in amplitude as the temperature is lowered.
More experimental and theoretical work is needed to clarify this issue.

In closing, we would like to comment on the possible physical origin of the
double resistance anomalies. Clearly the double resistance anomalies must
be due to similar physical process as the single resistance anomaly observed
previously in uniform mesoscopic superconducting samples. Unfortunately,
the physical origin of the single resistance anomaly is still subject
to an intensive debate
~\cite{santh91,moshc94,park97,strun98,aruty99,landa97,moshc97} and is not
settled. One school of thoughts has argued that charge imbalance
near the superconductor-normal metal (SN) interface contributes to the
resistance anomaly~\cite{moshc94,park97,strun98,moshc97} while another
school has pursued an alternative model based on the potential step at
the SN interface and the effects of the interface shape~\cite{aruty99,landa97}.
Given this, it is perhaps reasonable to delay this question regarding the
precise physical origin of the double resistance anomalies and the 
related excessive resistance to future inquiries.

We would like to acknowledge useful discussions with Dr. B. Pannetier.
This work is supported by NSF under grant DMR-0202534.

\end{document}